    \documentclass[prd,aps,preprint,tightenlines,showpacs,nofootinbib,superscriptaddress]{revtex4-1}
\usepackage{mathrsfs}
\usepackage{amsfonts}
\usepackage{amsmath}
\usepackage{amssymb}
\usepackage{array}
\usepackage{verbatim}
\usepackage{bm}
\usepackage{epsfig}
\usepackage{graphicx,color}
\usepackage{relsize}
\usepackage{lineno}
\usepackage{float}
\usepackage{multirow}
\RequirePackage{xspace}

\def\lsim{\raise0.3ex\hbox{$<$\kern-0.75em\raise-1.1ex\hbox{$\sim$}}}

\def\gsim{\raise0.3ex\hbox{$>$\kern-0.75em\raise-1.1ex\hbox{$\sim$}}}

\newcommand{\be}{\begin{equation}}

\newcommand{\ee}{\end{equation}}

\def\beq{\begin{equation}}

\def\eeq{\end{equation}}

\def\beqa{\begin{eqnarray}}

\def\eeqa{\end{eqnarray}}

\newcommand{\ba}{\begin{eqnarray}}

\newcommand{\ea}{\end{eqnarray}}

\def\gappeq{\mathrel{\rlap {\raise.5ex\hbox{$>$}}

{\lower.5ex\hbox{$\sim$}}}}

\def\lappeq{\mathrel{\rlap{\raise.5ex\hbox{$<$}}

{\lower.5ex\hbox{$\sim$}}}}

\def\Toprel#1\over#2{\mathrel{\mathop{#2}\limits^{#1}}}

\begin{document}

\title{Probing a $Z'$ gauge boson  via neutrino trident scattering in the Forward Physics Facility at the  LHC and FCC}

\vspace{3cm}

\author{Reinaldo {\sc Francener}}
\email{reinaldofrancener@gmail.com}
\affiliation{Instituto de Física Gleb Wataghin - Universidade Estadual de Campinas (UNICAMP), \\ 13083-859, Campinas, SP, Brazil. }

\author{Victor P. {\sc Gon\c{c}alves}}
\email{barros@ufpel.edu.br}
\affiliation{Institute of Physics and Mathematics, Federal University of Pelotas, \\
  Postal Code 354,  96010-900, Pelotas, RS, Brazil}
%\affiliation{Institute of Modern Physics, Chinese Academy of Sciences,
%  Lanzhou 730000, China}

\author{Diego R. {\sc Gratieri}}
\email{drgratieri@id.uff.br}
\affiliation{Escola de Engenharia Industrial Metal\'urgica de Volta Redonda,
Universidade Federal Fluminense (UFF),\\
 CEP 27255-125, Volta Redonda, RJ, Brazil}
\affiliation{Instituto de Física Gleb Wataghin - Universidade Estadual de Campinas (UNICAMP), \\ 13083-859, Campinas, SP, Brazil. }

\vspace{3cm}

\begin{abstract}
The study of neutrino physics at the Large Hadron Collider is already a reality, and a broad neutrino physics program is expected to be developed in forthcoming years at the Forward Physics Facility (FPF). In particular, the neutrino trident scattering process, which is a rare Standard Model process, is expected to be observed for the first time with a statistical significance of $5\sigma$ using the FASER$\nu$2 detector. { Moreover, similar studies are expected to be performed in the proposed Future Circular Collider (FCC).}  Such perspectives motivate the investigation of the impact of New Physics on the predictions for the corresponding number of events. In this letter, we consider the $L_\mu - L_\tau$ model, which  predicts an additional massive neutral gauge boson, $Z'$, that couples to neutrino and charged leptons of the second and third families, and estimate the production of a dimuon system in the neutrino trident scattering at the FASER$\nu$2 assuming different models for the incoming neutrino flux at the LHC and FCC energies. { We derive the associated sensitivity and demonstrate that the FPF@LHC is not able to improve the current bounds on the $L_\mu - L_\tau$ model, while a future measurement of the dimuons produced in neutrino trident events in the FPF@FCC will extend the coverage of the parameter space in comparison to previous experiments.}
\end{abstract}


\keywords{Neutrino tridents; Z' boson; Forward Physics Facility.}

\maketitle

\vspace{1cm}

%\section{Introduction}

\begin{figure}[b]
	\centering
	\begin{tabular}{ccccc}
\includegraphics[width=0.5\textwidth]{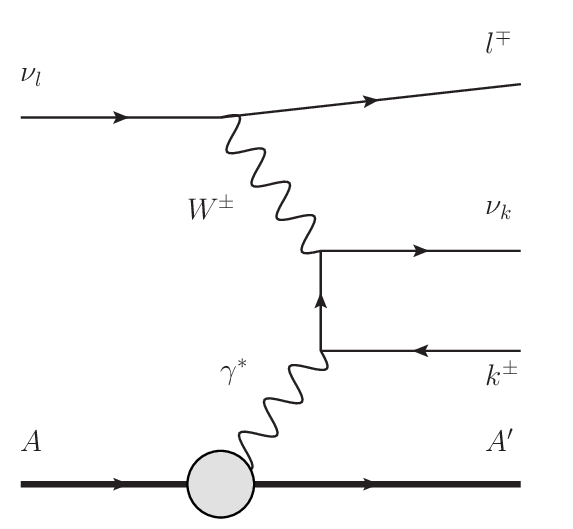} &
\includegraphics[width=0.5\textwidth]{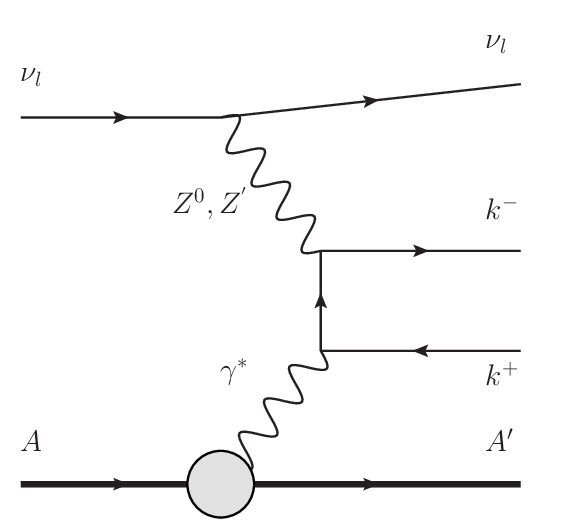} 
	\end{tabular}
\caption{ Feynman diagrams for the neutrino trident scattering off a nucleus target associated with a $W^\pm$ (left), $Z^0$ and $Z'$ (right) exchange. The lepton flavors $l$ and $k$ can be equal or different, and $A^{\prime} = A$ ($A^{\prime} \neq A$) in coherent (incoherent) interactions. }
\label{fig:diagrams}
\end{figure}

The recent results obtained by the FASER \cite{FASER:2021mtu,FASER:2023zcr} and SND@LHC \cite{SNDLHC:2023pun} experiments have demonstrated that the Large Hadron Collider (LHC) can also be considered a neutrino factory, which allow developing a neutrino physics program complementary to those explored by the IceCube Observatory and by current and future facilities, such as the DUNE - near detector \cite{DUNE:2020lwj}. This program is expected to be significantly extended during the  high - luminosity phase of the LHC (HL-LHC), with the construction and operation of the Forward Physics Facility (FPF) \cite{Anchordoqui:2021ghd,Feng:2022inv}, which will house a suite of new dedicated experiments focused on improve our understanding about neutrino physics at TeV energies and search for Beyond of the Standard Model (BSM) signals. { Moreover, the results recently derived in Ref. \cite{MammenAbraham:2024gun}   indicated that a far - forward experiment integrated within the proposed Future Circular Collider (FCC) project, denoted FPF@FCC hereafter, can increase by several orders of magnitude the flux of neutrinos and number of events available to physical studies compared to FPF@LHC.} Such  expectations have motivated an intense phenomenology over the last years (See, e.g. Refs. \cite{Bai:2020ukz,Bai:2021ira,Bai:2022xad,Buonocore:2023kna,Bhattacharya:2023zei,Maciula:2022lzk,Aboubrahim:2022qln,Cheung:2023gwm,Cruz-Martinez:2023sdv,Feng:2024zgp,Anchordoqui:2024ynb}). In particular, the studies performed in Refs. \cite{Francener:2024wul,Altmannshofer:2024hqd}    indicated that the neutrino trident scattering, which is a rare SM process, could be measured for the first time at the LHC in the forthcoming years.
This process is characterized by the production of a  pair of charged leptons by the neutrino scattering off the Coulomb field of a heavy nucleus. In the Standard Model, such a weak process occurs through a $W^{\pm}$ or a $Z^0$ boson exchange, as represented in Fig. \ref{fig:diagrams}. If observed, the neutrino trident events are expected to provide important constraints in several BSM scenarios, as demonstrated in the studies performed e.g. in Refs. \cite{Altmannshofer:2014pba,Ge:2017poy,Magill:2017mps,Altmannshofer:2019zhy,Ballett:2019xoj,Bigaran:2024zxk,Altmannshofer:2024hqd}.

In this letter, we will focus on the $L_\mu - L_\tau$ model \cite{He:1990pn,He:1991qd}, which is based on gauging the difference between the $\mu$ - lepton number and the $\tau$ - lepton number, and  predicts an additional massive neutral gauge boson, $Z'$, that couples to neutrino and charged leptons of the second and third families. The associated  Lagrangian is given by 
\begin{eqnarray}
    \mathcal{L} = g' Z_{\alpha}' [
    (\bar{\mu} \gamma^{\alpha} \mu) - 
    (\bar{\tau} \gamma^{\alpha} \tau) + 
    (\bar{\nu}_{\mu} \gamma^{\alpha} P_{L} \nu_{\mu}) - 
    (\bar{\nu}_{\tau} \gamma^{\alpha} P_{L} \nu_{\tau})
    ] ,
    \label{eq:lagrangian}
\end{eqnarray}
where $g'$ is the gauge coupling and $P_L$ is the chirality projector. Over the last years, the implications of this model have been largely investigated, and several experimental studies have restricted the values of the gauge coupling $g'$ and its  mass $m_{Z'}$. In particular, the results derived in Refs. \cite{Altmannshofer:2014pba,Ballett:2019xoj}  indicated that the study of the trident scattering process is one of the most promising ways to constraint a  large region of the parameter space.  This model implies that the trident process, at the lowest order,  can also occur through a $Z'$ exchange, as represented in the right panel of Fig. \ref{fig:diagrams} for an incoming muon or tau neutrino. In what follows, we will investigate the impact of a $Z'$ gauge boson, as described by the $L_\mu - L_\tau$ model,  on trident scattering events with two muons in the final state, produced in neutrino - tungsten interactions at the FPF during the { HL - LHC and FCC eras}. As in Ref. \cite{Francener:2024wul}, our results { for the FPF@LHC} will be derived considering the characteristics expected for the FASER$\nu$2 detector \cite{Anchordoqui:2021ghd,Feng:2022inv}. { For the FPF@FCC, we will consider the three different options of neutrino detector considered in Ref. \cite{MammenAbraham:2024gun}, which taken the geometry and kinematic acceptance of  FASER$\nu$2 as a baseline.  }

{ Initially, let's perform the analysis for the FPF@LHC}. The number of trident scattering events in the FASER$\nu$2 detector  can be obtained by the product of the neutrino flux integrated over time with the scattering probability, given by $\sigma_{\nu A} \rho L /m_{A}$, where $\rho$ is the density of the detector, $L$ its length and $m_A$ the mass of the target nucleus  (tungsten for FASER$\nu$2). The main ingredient for calculating the interaction probability is the neutrino-nucleus cross-section for the trident scattering. As in Ref. \cite{Francener:2024wul}, the SM contributions for this cross-section will be estimated  using the effective model of contact interaction between the four leptons involved in the scattering \cite{Altmannshofer:2019zhy}. This approximation can be used in the energy regime of interest, given that the square of the four-momentum transferred by the neutrino is expected to be much smaller than the mass of the bosons $W^{\pm}$ and $Z^{0}$. For the $L_{\mu} - L_{\tau}$ model,  we will follow the implementation performed in Ref. \cite{Altmannshofer:2019zhy}   and  do not assume this approximation, since we will consider the mass of $Z'$ in a range of $10^{-3} - 10 $ GeV. Our results will be  for the $2 \rightarrow 4$ calculation, without assuming the validity of the equivalent photon approximation. Moreover, we will add the  coherent  and incoherent contributions for the trident process~\footnote{In coherent processes,  the scattering occurs with the full target nucleus, which remains intact. In contrast, in incoherent processes, the interaction occurs with the individual nucleons of the nucleus. For a more detailed discussion see, e.g., Ref. \cite{Ballett:2018uuc}.}.  Finally,  as in Ref. \cite{Francener:2024wul}, we will assume  a Woods-Saxon distribution for the nuclear electric charge \cite{Woods:1954zz}, with parametrization provided in \cite{DeVries:1987atn}.

Another ingredient needed to estimate the number of neutrino trident events at the FASER$\nu$2 is the time integrated neutrino flux. For this quantity, we will use the expected neutrino fluxes for the HL - LHC era obtained in Ref. \cite{Kling:2023tgr} considering different hadronic models for the treatment of the meson production at ultra - forward rapidities in  $pp$ collisions at the ATLAS detector, and assuming $\sqrt{s_{pp}} = 14$ TeV.  We will  consider the predictions associated with these different  models implemented in different Monte Carlo generators to estimate the current theoretical uncertainties in the predictions for the neutrino trident events at the FASER$\nu$2. For the flux of neutrinos arising from the decay of light mesons, we will consider the predictions of the   \texttt{EPOS-LHC} \cite{Pierog:2013ria}, \texttt{DPMJET 3.2019.1} \cite{Roesler:2000he,Fedynitch:2015}, \texttt{QGSJET II-04} \cite{Ostapchenko:2010vb}, \texttt{SIBYLL 2.3d} \cite{Ahn:2009wx,Riehn:2015oba} and \texttt{Pythia 8.2} \cite{Fieg:2023kld,Sjostrand:2014zea} Monte Carlo generators. On the other hand, to describe the component of the neutrino flux arising from heavy mesons, we will use the results derived in the references  \cite{Bai:2020ukz,Bai:2021ira,Bai:2022xad} (denoted \texttt{BDGJKR}),  \cite{Buonocore:2023kna} (denoted \texttt{BKRS}),  \cite{Bhattacharya:2023zei} (denoted \texttt{BKSS} $k_T$), \cite{Maciula:2022lzk}  (denoted \texttt{MS} $k_T$)  and  \cite{Ahn:2011wt,Fedynitch:2018cbl} (denoted \texttt{SIBYLL 2.3d}). The total flux of neutrinos will be the combination of components arising from the decays of light mesons and heavy mesons. For the SM events, we will consider the contributions of the three flavors of neutrinos, with electronic neutrinos only contributing to two muons in the final state through the exchange of the $Z^{0}$ boson. In contrast, for the events induced by the exchange of the $Z'$ boson, we will only consider the fluxes of muonic and tauonic neutrinos, given that in the $L_\mu - L_\tau$ model, this boson does not couple with leptons from the electron family.

As the FASER$\nu$2 will have some limitations for detecting trident scattering, we will consider the following experimental cuts in our results: ($i$) we are considering the efficiency in the reconstructing the two muons given in \cite{Altmannshofer:2024hqd}, { which considers an efficiency $\leq 1$ for muons with energy less than $30$ GeV}; ($ii$) an angular separation between the two muons ($\theta$) less than 0.1 rad; and ($iii$) for the contribution of the incoherent process, we will impose that  the energy of the proton in the final state must be less than 1.2 GeV, in order to characterize the absence of hadronic activity in the scattering. It is important to emphasize that there are other processes that can generate two muons in the final state, such as charged current interaction of muon neutrinos that give rise to heavy mesons that quickly decay into muons, or the production of energetic charged pions that can be mistakenly identified as muons. The results presented in Ref. \cite{Altmannshofer:2024hqd} indicated that these backgrounds can be strongly suppressed by applying cuts in the  multiplicity of charged particles in the final state, in the distance between the production vertex of the two muons, in the angular separation between the two muons and using muon identification to disregard pions and energetic kaons of the final state. Considering this perspective, in this work, we will assume that the dimuons associated with the trident events can be separated with negligible background. { The impact of the kinematic cuts on the number of events, derived considering the different models for the incoming neutrino flux, is presented in Table \ref{table:Nevents}. One has that the cut on the angular separation reduces the number of events by $\approx 10$ \%. On the other hand, the cut on the muon energy, implies a reduction of $\approx 40$ \% in the number of events. The cut on nucleon energy in the final state for incoherent interactions does not result in an appreciable change in the number of events.  Our results indicate that the predicted number of events at the FASER$\nu$2, after the inclusion of the kinematical cuts, is larger than 17 events.}

\begin{center}
	\begin{table}[t]
		\begin{tabular}{|c|c|c|c|c|}
%			\hline
			\hline 
			& Without cuts & Angular separation  &  $\theta \leq 0.1$ rad + muon energy cut \\
            		&  &  $\theta \leq 0.1$ rad &   \\
%			\hline 	
			\hline
Pythia8 + MS $k_T$ 	    & 36.3 & 32.4 & 19.6
\tabularnewline
%			\hline 
DPMJET + BKRS 		    & 34.0 & 30.4 & 18.7
\tabularnewline
%			\hline 
EPOS-LHC + BKSS $k_T$ 	& 54.2 & 48.7 & 30.8
\tabularnewline
%			\hline 
QGSJET + BDGJKR 	    & 33.8 & 29.9 & 17.4
\tabularnewline
%			\hline 
SIBYLL 		       	    & 32.4 & 29.0 & 17.8
\tabularnewline
%			\hline 
            \hline
		\end{tabular}
		\caption{ Number of trident events in the FASER$\nu$2 detector for muon pairs in the final state during the high luminosity era of the LHC, estimated considering different MC generators for the incident neutrino flux. We present both the total expected and the signal after experimental cuts. }
		\label{table:Nevents}
	\end{table}
\end{center}

\begin{figure}[t]
	\centering
	\begin{tabular}{ccccc}
\includegraphics[width=0.8\textwidth]{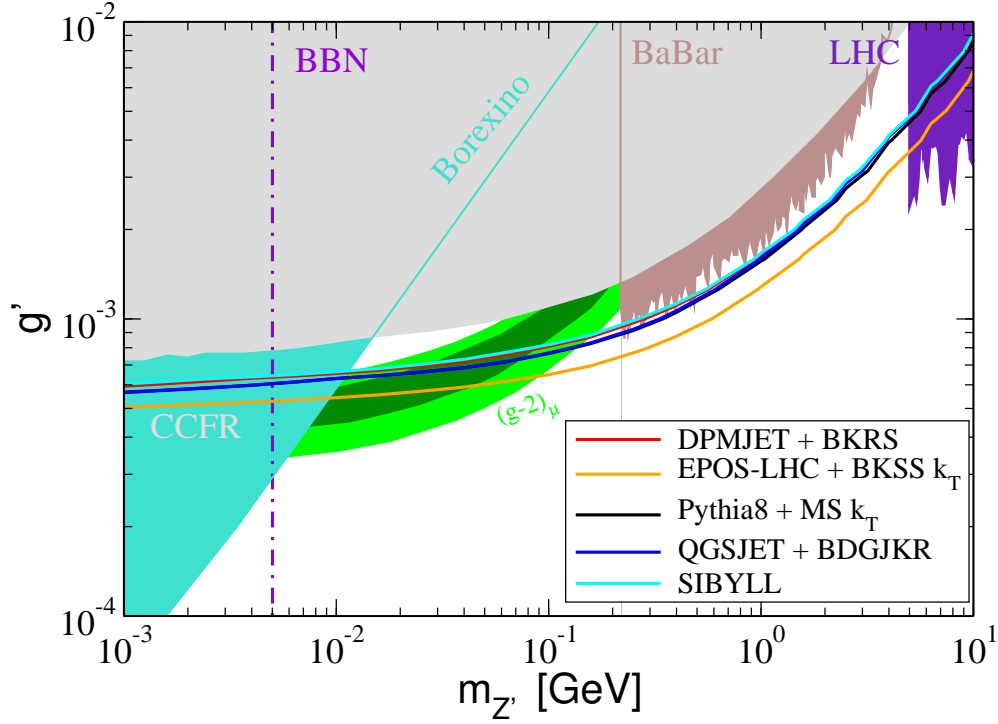} 
	\end{tabular}
\caption{Sensitivity of the { far - forward detectors at the LHC (FASER$\nu$2) and FCC} for the neutrino trident events associated with the $Z'$ gauge boson predicted by the $L_\mu - L_\tau$ model. Results,  at the 2$\sigma$ level, derived considering different MC generators for the incoming neutrino flux.  For comparison, the  existing constraints from other processes and experiments are also presented. The green bands represent the parameter space in which the presence of a $Z'$ solves the muon magnetic moment anomaly at the $1\sigma$ and $2\sigma$ levels. The LHC band represents the region excluded  in $pp$ collisions at the LHC, considering the analysis performed by the CMS \cite{CMS:2018yxg} and ATLAS \cite{ATLAS:2023vxg,ATLAS:2024uvu} collaborations of the production of $Z^{0}$ and $W^{\pm}$ bosons and subsequent decay in the $Z^{0} \rightarrow Z' \mu^{\pm} \mu^{\mp} \rightarrow \mu^{\pm} \mu^{\mp} \mu^{\pm} \mu^{\mp}$ and  $W^{\pm} \rightarrow Z' \mu^{\pm} \nu_{\mu} \rightarrow \mu^{\pm} \mu^{\mp} \mu^{\pm} \nu_{\mu}$ channels. The BaBar band represents the associated excluded region in the study of the  $Z^{0}$ boson decay process \cite{BaBar:2016sci}. We also have in the figure the excluded regions by the Borexino experiment in interactions of neutrinos with electrons \cite{Bellini:2011rx,Harnik:2012ni}, which was constructed considering that the interaction rate should not be more than $8\%$ greater than the SM prediction. {The magenta dotted line represents the recent results from the NA64$\mu$ experiment with 90\% CL \cite{NA64:2024klw}}. The bounds from Big Bang Nucleosynthesis (BBN) derived in  \cite{Escudero:2019gzq} and from Planck 2018 \cite{Planck:2018vyg,Ghosh:2024cxi}, which implies that $m_{Z'} \gtrsim 5 \,\mathrm{MeV}$ and $m_{Z'} \gtrsim 10 \,\mathrm{MeV}$, respectively, are also presented. Finally, the excluded region derived by the CCFR experiment in the neutrino trident process with two muons in the final state \cite{CCFR:1991lpl,Altmannshofer:2014pba} is also presented. }
\label{fig:mapa1}
\end{figure}

In order to estimate the impact of a new gauge boson in the neutrino trident events, we will perform an analysis of $\chi^2$ for different values of the model parameters $g'$ and $M_{Z'}$ using Poisson statistics, which is justified given the low number of trident events expected in FASER$\nu$2 for muon pairs in the final state { (See Table \ref{table:Nevents})}. As a consequence,  $\Delta \chi^{2}$ will be given by \cite{ParticleDataGroup:2024cfk,Algeri:2019lah}
\begin{eqnarray}
\Delta \chi ^{2} = 
2
%\sum_{i=1}^{n}
\left[ (N^{L_{\mu} - L_{\tau}} - N^{SM}) +  N^{SM}\mathrm{ln}\left( \frac{N^{SM}}{ N^{L_{\mu} - L_{\tau}}} \right) \right] \, ,
    \label{eq:deltaChi2}
\end{eqnarray}
where $N$ is the total number of trident scattering events expected after all experimental cuts.
%where $N$ is the number of trident scattering events expected in the { incident neutrino} energy bin $i$ for each of the analyzed models. { In our analysis, we are considering the binning given in the neutrino flux obtained in ref. \cite{Kling:2023tgr}. Such a binning procedure is important since that in the  considered range for the $Z'$ mass ($m_{Z^{\prime}} < m_{Z^0},\, m_W$),  the energy dependence of the trident and SM cross-sections is different, mainly at smaller neutrino energies.}

%The $Z'$ contribution in its analyzed mass range shows a different cross-section shape compared to the SM bosons' contribution to the events, providing a greater contribution at low neutrino energy, due to its smaller mass relative to the $W^{\pm}$ and $Z^{0}$ bosons. This implies that it is important to consider binned events.}  

In Fig. \ref{fig:mapa1} we present our results for the sensitivity of FASER$\nu$2 detector to the presence of a new gauge boson $Z'$ in the trident process with two muons in the final state, derived considering 
different Monte Carlo generators for the incoming neutrino flux and considering the significance of 2$\sigma$. For comparison, we also present the existing excluded regions for the mass and coupling parameter space ($m_{Z'}, g'$)  derived using the experimental data from BaBar\cite{BaBar:2016sci},  LHC\cite{CMS:2018yxg,ATLAS:2023vxg,ATLAS:2024uvu}, CCFR\cite{CCFR:1991lpl,Altmannshofer:2014pba}, Borexino\cite{Bellini:2011rx,Harnik:2012ni}, NA64$\mu$\cite{NA64:2024klw} and Planck\cite{Planck:2018vyg} experiments. Moreover, we also present in the green bands,  the regions allowed at 1$\sigma$ and 2$\sigma$ level, in which the presence of  a $Z'$ solve the anomaly observed in the $(g-2)_{\mu}$ experiment \cite{Keshavarzi:2018mgv}. { It is important to emphasize that recent results from lattice quantum chromodynamics (QCD) for the hadronic vacuum polarization contribution to the $(g-2)_{\mu}$ (See, e.g., Refs. \cite{
Boccaletti:2024guq, Djukanovic:2024cmq, FermilabLattice:2024yho,Bazavov:2024eou}) have decreased the tension between the SM predictions and  the experimental data, implying a reduction of the region in the parameter space where the $Z^{\prime}$ solves the anomaly. } 
%One has that the  predictions derived considering the different models are similar, unless for the \texttt{EPOS-LHC} + \texttt{BKSS} $k_{T}$ prediction, which implies that the FASER$\nu$2 detector is able to probe a wider region of the parameter space in comparison to the other hadronic models. Such behavior is  due to the larger total neutrino events predicted by an incoming neutrino flux derived using the \texttt{EPOS-LHC} + \texttt{BKSS} $k_{T}$ model (See Fig. 5 in Ref. \cite{Francener:2024wul}). 
Our results using different MC generators form a narrow black band in the exclusion limit, caused by the difference in the number of events predicted for each neutrino flux model (See Fig. 5 in Ref. \cite{Francener:2024wul}).
One has that for the $m_{Z'} \gtrsim 5$ GeV region, $g'$ is proportional to $m_{Z'}$. This happens because in this region, the mass of $Z'$ becomes much greater than the square four-momentum transferred by the neutrino, and the propagator becomes proportional to $1/m_{Z'}^{2}$. Conversely, for the region of $m_{Z'} \lesssim 10^{-2}$ GeV there is a saturation, because in this case the mass of $Z'$ becomes negligible compared to the square four-momentum transferred by the neutrino incident. { The results for the FPF@LHC presented in Fig. \ref{fig:mapa1} indicate that, independently of the hadronic model considered, the FASER$\nu$2 experiment will not be able to improve the current bounds in the $L_\mu - L_\tau$ model.}

%some regions of the parameter space not yet covered by previous experiments, including a part of the region where the $Z'$ can explain the $(g-2)_{\mu}$ data. In partiOur results indicate that  FASER$\nu$2 will cover

%\begin{figure}[t]
%	\centering
%	\begin{tabular}{ccccc}
%\includegraphics[width=0.8\textwidth]{mapa_MZP_GP_bandas.eps} 
%	\end{tabular}
%\caption{Predictions for the sensitivity at the 1$\sigma$ and 2$\sigma$ levels of the  FASER$\nu$2 detector for the neutrino trident events associated with the $Z'$ gauge boson predicted by the $L_\mu - L_\tau$ model. Results derived considering the current theoretical uncertainty on the incoming neutrino flux. The  existing constraints from other processes and experiments are also presented.}
%\label{fig:mapa2}
%\end{figure}

%Finally, in Fig. \ref{fig:mapa2} we again present the  $Z'$ parameter space, but now with our predictions for 1$\sigma$ and 2$\sigma$ sensitivity levels at FASER$\nu$2, with the uncertainty bands  constructed considering the different MC generators for the incident neutrino flux. 

\begin{center}
	\begin{table}[t]
		\begin{tabular}{|c|c|c|c|c|}
%			\hline
			\hline 
			& Without cuts & Angular separation  &  $\theta \leq 0.1$ rad + muon energy cut \\
            		&  &  $\theta \leq 0.1$ rad &   \\
%			\hline 	
			\hline
FCC$\nu$                & 6305.5 & 5799.3 & 4078.8
\tabularnewline
FCC$\nu$(w)             & 19006.9 & 17363.8 & 11840.4
\tabularnewline
FCC$\nu$(d)             & 63220.6 & 58148.6 & 40907.4
\tabularnewline
            \hline
		\end{tabular}
		\caption{ Number of trident events at FPF@FCC for muon pairs in the final state, estimated considering the EPOS-LHC for light meson and the POWHEG matched with Pythia8.3 for heavy meson contribution for the incident neutrino flux. We present both the total expected and the signal after experimental cuts. }
		\label{table:Nevents_FCC}
	\end{table}
\end{center}

{ Considering  the perspective of a proton - proton collision program in the proposed Future Circular Collider, characterized by larger center - of - mass energies and higher integrated luminosities, which could record up to a factor 1000 more neutrino events compared to its LHC counterpart \cite{MammenAbraham:2024gun}, it is natural to investigate if a FPF@FCC could be able to enlarge the bounds in the $L_\mu - L_\tau$ model. In what follows, we extend the analysis performed above for this proposed experiment. We will consider proton-proton collisions with a center of mass energy of 100 TeV and an integrated luminosity of 30 ab$^{-1}$. In our calculations, we will use the  neutrino fluxes obtained in Ref. \cite{MammenAbraham:2024gun}, which considers a neutrino detector installed 1500 m from the interaction point. The neutrino flux coming from light  hadrons has been obtained using the \texttt{EPOS-LHC} \cite{Pierog:2013ria} Monte Carlo, while that associated with the decay of heavy mesons was obtained using the \texttt{POWHEG} MC \cite{Alioli:2010xd} matched with \texttt{Pythia8.3} \cite{Bierlich:2022pfr} generator. As in Ref. \cite{MammenAbraham:2024gun}, we will assume the same detector target (tungsten) and the same experimental cuts described above for FASER$\nu$2, plus an additional cut in the incident neutrino energy. Our results are for incident neutrinos up to 5 TeV, given that for more energetic neutrinos an additional background from $W^{\pm}$ boson production would arise \cite{Zhou:2019vxt}. Finally, we will also consider three different geometries for the detector: (a) with the same dimensions of the FASER$\nu$2 (FCC$\nu$), considered as a baseline, and with two possible improvements in size, (b) one with transverse area 10 times larger that of the baseline detector [FCC$\nu$(w)] and (c) another with a length 10 times that of the baseline detector [FCC$\nu$(d)]. In Table \ref{table:Nevents_FCC} we present the results for trident events with two muons in the final state for experiments FASER$\nu$2 like in the proposed FCC. Our results for the trident scattering in the FPF@FCC, presented here for the first time, indicate an increasing between 100 and 1000 times in comparison to the number of trident events expected at FASER$\nu$2 in the HL-LHC. Regarding the impact of this increasing on the sensitivity to a new $Z^{\prime}$ gauge boson, the corresponding results are presented  in Fig \ref{fig:mapa1}. One has that  for all three possible experiments analyzed,  new regions in the parameter space  will be covered, both in the $Z'$ region of lower masses ($\approx 1$ GeV), and for masses greater than that of the electroweak gauge bosons ($W^{\pm}$ and $Z^{0}$). In particular, for the FCC$\nu$ detector, it will be able to probe the  mass ranges of $0.3 \le m_{Z^{\prime}} \le 4$ GeV and $m_{Z^{\prime}} \ge 50$ GeV, which are not covered by current experiments.  Such regions are enlarged considering the two other options for the neutrino detector, especially in the FCC$\nu$(d) case.}
Finally, it is important to emphasize that the results presented in Fig. \ref{fig:mapa1} were derived considering only statistical uncertainties and that the inclusion of systematic uncertainties is expected to slightly reduce the FCC sensitivity.
%related to the neutrino flux, neutrino-target cross-section, and detector efficiency are expected to become significant. Consequently, a more realistic sensitivity estimate for the $Z'$ parameter space may be slightly reduced compared to the result presented in Fig. \ref{fig:mapa1}. 

{ As a summary, in this letter we have investigated the potentiality of the Forward Physics Facility to searching for a new gauge boson $Z'$ in neutrino trident events in the FASER$\nu$2 detector at the LHC and possible far - forward experiments at the proposed FCC. We have estimated the contribution associated with the $Z'$ exchange predicted by the $L_\mu - L_\tau$ model and derived the sensitivity to new physics considering different values for the gauge coupling and $Z'$ mass. We have obtained that the FPF@LHC will be not able to improve the current bounds. In contrast, our results for the FPF@FCC indicated that a future measurement of the dimuons produced in neutrino trident events at FCC will probe an uncovered region of the parameter space, allowing to test the predictions of the $L_\mu - L_\tau$ model.
}

\begin{acknowledgments}
{ We would like to thank the anonymous referee for many
useful suggestions and comments which have helped us
improve our manuscript.}  R. F. acknowledges support from the Conselho Nacional de Desenvolvimento Cient\'{\i}fico e Tecnol\'ogico (CNPq, Brazil), Grant No. 161770/2022-3. V.P.G. was partially supported by CNPq, FAPERGS and INCT-FNA (Process No. 464898/2014-5). D.R.G. was partially supported by CNPq and MCTI.

\end{acknowledgments}

\hspace{1.0cm}

\end{document}